\documentclass[11pt,showpacs,preprintnumbers,amsmath,amssymb,prd,nofootinbib,superscriptaddress]{revtex4-2}

\usepackage{dcolumn}% Align table columns on decimal point
\usepackage{bm}% bold math
\usepackage{ifpdf}
\usepackage{hyperref}
\usepackage{xcolor,color,graphicx,graphics,physics}
\usepackage[spanish,english]{babel}%, portuguese
\usepackage[latin1]{inputenc}
\usepackage[OT1]{fontenc}
\usepackage{latexsym,amssymb,amsmath,amsfonts, slashed,cancel, simpler-wick}
\usepackage{makeidx}
\usepackage{epsfig,subfigure}
\usepackage{natbib}
\usepackage{epstopdf}
\usepackage{mathrsfs}
\usepackage{hyperref}%\usepackage[colorlinks=true,linkcolor=blue]{hyperref}
\hypersetup{colorlinks=true, linkcolor=blue, citecolor=blue, urlcolor=blue}
\usepackage{enumerate}
\usepackage{tikz}
\usepackage{feynmp-auto}  % Alternativa para diagramas
\usepackage{tikz-feynman}
\tikzfeynmanset{compat=1.1.0}
\usepackage{fixmath}

\everymath{\displaystyle}
\usepackage{graphicx}

\usepackage[T1]{fontenc}
\usepackage{amsmath}
\usepackage{amssymb}
\usepackage{graphicx}
\usepackage{xcolor}

\newcommand{\bea}{\begin{eqnarray}}
\newcommand{\eea}{\end{eqnarray}}

\newcommand{\orcid}[1]{\href{https://orcid.org/#1}{\includegraphics[width=10pt]{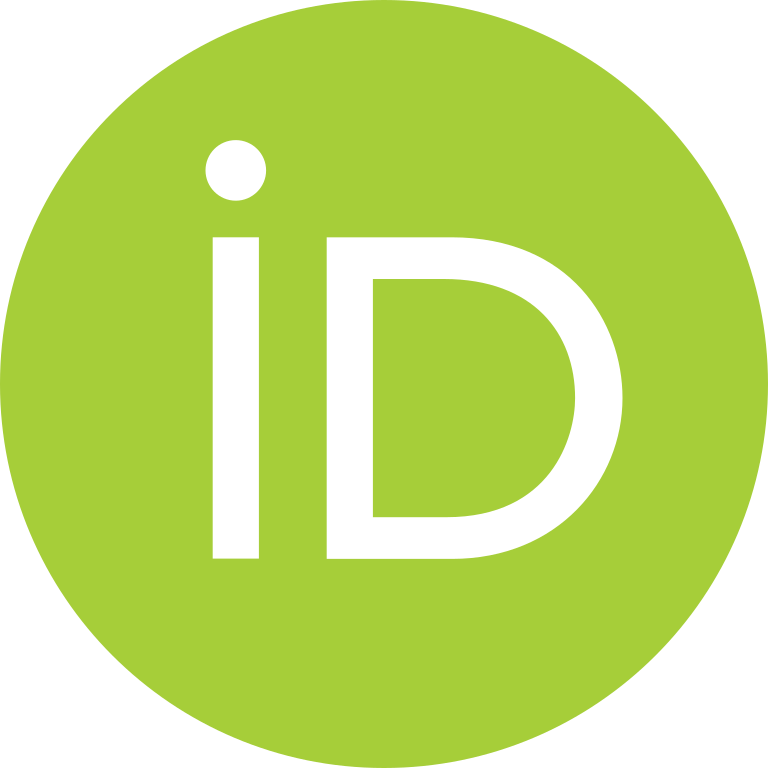}}}

\begin{document}

%\title{Lorentz-violating scalar theory at finite temperature}
\title{Thermal and Casimir effects in a Lorentz-violating massive scalar field}

\author{D. S. Cabral  \orcid{0000-0002-7086-5582}}
\email{danielcabral@fisica.ufmt.br}
\affiliation{Programa de P\'{o}s-Gradua\c{c}\~{a}o em F\'{\i}sica, Instituto de F\'{\i}sica,\\ 
Universidade Federal de Mato Grosso, Cuiab\'{a}, Brasil}

\author{L. H. A. R. Ferreira \orcid{0000-0002-4384-2545}}
\email{luiz.ferreira@fisica.ufmt.br}
\affiliation{Programa de P\'{o}s-Gradua\c{c}\~{a}o em F\'{\i}sica, Instituto de F\'{\i}sica,\\ 
Universidade Federal de Mato Grosso, Cuiab\'{a}, Brasil}

\author{L. A. S. Evangelista \orcid{0009-0002-3136-2234}}
\email{lucassouza@fisica.ufmt.br}
\affiliation{Programa de P\'{o}s-Gradua\c{c}\~{a}o em F\'{\i}sica, Instituto de F\'{\i}sica,\\ 
	Universidade Federal de Mato Grosso, Cuiab\'{a}, Brasil}

\author{A. F. Santos \orcid{0000-0002-2505-5273}}
\email{alesandroferreira@fisica.ufmt.br}
\affiliation{Programa de P\'{o}s-Gradua\c{c}\~{a}o em F\'{\i}sica, Instituto de F\'{\i}sica,\\ 
Universidade Federal de Mato Grosso, Cuiab\'{a}, Brasil}

\begin{abstract}

In this work, a massive scalar field theory incorporating Lorentz violation is investigated. The symmetry breaking is introduced via a background traceless antisymmetric tensor. Within the framework of Thermo Field Dynamics (TFD), the effects of space-time compactification are explored, allowing the simultaneous treatment of thermal and finite-size phenomena. The resulting modifications to the energy-momentum tensor and Feynman propagator are analyzed, leading to Lorentz-violating corrections to the Stefan-Boltzmann law and the Casimir effect. This unified approach highlights the interplay between temperature, spatial constraints, and Lorentz-violating backgrounds in shaping the behavior of quantum fields.

\end{abstract}

\maketitle

\section{Introduction}

With the advent of modern physics, quantum mechanics introduced new approaches to understanding physical phenomena. However, these advances also revealed challenges, particularly in quantum field theory, where divergent zero-point energies naturally arise from canonical quantization \cite{plunien1986casimir}. Typically, these divergences are handled by normal ordering of operators, which sets the vacuum expectation value of the Hamiltonian to zero. Yet, since vacuum fluctuations affect the universe's energy density, these divergences may lead to observable physical consequences. Therefore, properly addressing and interpreting vacuum energy is essential for understanding fundamental quantum systems.

Building on these foundational ideas, H. Casimir proposed in 1948 \cite{Casimir:1948dh} that two parallel conducting plates placed in a quantum vacuum experience an attractive force. This prediction was experimentally confirmed nearly a decade later \cite{sparnaay1958measurements}, establishing the Casimir effect as a striking manifestation of quantum mechanics. The attraction arises from boundary conditions or topological effects imposed on quantum fields by the plates, which alter the vacuum energy of the system. Since then, extensive research has explored the Casimir effect in various settings \cite{mehra1967temperature, schwinger1978casimir, lambrecht2006casimir, plunien2000dynamical, rodriguez2011casimir, johansson2009dynamical}, including gravitational backgrounds \cite{bezerra2017casimir, quach2015gravitational, hu2017gravitational, sorge2005casimir, sorge2019casimir, lambiase2017casimir}. Moreover, the effect has been generalized to include thermal corrections through Thermo Field Dynamics (TFD), a framework that introduces temperature as a compactification of the temporal dimension.

TFD is a real-time formalism in quantum field theory that incorporates thermal effects in both equilibrium and non-equilibrium systems \cite{Umezawa1, Umezawa2, Umezawa22, arimitsu1987non, Book, Khanna1, Khanna2, Santana1, Santana2}. It establishes a correspondence between the statistical average and the vacuum expectation value in quantum field theory. Numerous studies have applied TFD to diverse physical processes \cite{santos2016gravitational, santos2021casimir, cabral2025non, cabral2023compton, borrelli2021finite, abdalla2005closed, hashizume2013understanding, santana1996thermal, santos2022thermal, evangelista2025gravitational}. The formalism relies fundamentally on two key elements: the doubling of the Hilbert space and the Bogoliubov transformation. The doubled Hilbert space $\mathcal{H}_T$ is constructed as the tensor product of the original Hilbert space $\mathcal{H}$ (the non-tilde space) and an identical tilde space $\tilde{\mathcal{H}}$, i.e., $\mathcal{H}_T = \mathcal{H} \otimes \tilde{\mathcal{H}}$. These two spaces are related by the tilde conjugation rules \cite{Book}, which associate each operator in the original space with two corresponding operators in the doubled space. The Bogoliubov transformation then implements a rotation mixing the tilde and non-tilde operators, introducing thermal effects at the operator level. Although alternative formalisms exist for introducing temperature into quantum systems - such as the imaginary-time Matsubara approach \cite{matsubara} and the real-time Closed-Time Path (CTP) formalism \cite{schwinger1961brownian} - TFD proves particularly effective in the present context due to its topological formulation. In TFD, the space-time topology is expressed as $\Gamma^1_D = (\mathbb{S}^1)^\varrho \times \mathbb{R}^{D-\varrho}$, where $D$ denotes the space-time dimensionality and $\varrho$ is the number of compactified dimensions. This structure enables the compactification of any dimension into a hypertorus, represented by $(\mathbb{S}^1)^\varrho$, thereby allowing both thermal and spatial effects to be encoded through compactification parameters.

In this work, we investigate a massive scalar field in the presence of a Lorentz-violating background, considering the following topological configurations: (i) compactification of the time coordinate into a circle of circumference $\beta$, with $\beta$ being the inverse temperature; (ii) compactification of the spatial coordinate $z$ into a circle of length $L$; and (iii) simultaneous compactification of both coordinates. These settings offer a unified framework for describing distinct physical phenomena through a common compactification approach. In particular, we analyze the Stefan-Boltzmann-type law and the Casimir effect, both at zero and finite temperature. Since these effects naturally arise from coordinate compactification, the analysis underscores the versatility and effectiveness of TFD in incorporating thermal and finite-size corrections into quantum field theory. Furthermore, we extend our analysis to the context of the Standard-Model Extension (SME), focusing on the influence of a Lorentz-violating background. In this framework, symmetry-breaking effects may significantly impact the physical processes under study.

The Standard Model Extension (SME) is a widely studied framework in particle physics that encompasses numerous approaches to extending the well-established Standard Model. One prominent direction within the SME involves introducing coupling terms in the Lagrangian that explicitly break Lorentz symmetry. Since the late twentieth century, the effects of Lorentz violation (LV) have been extensively investigated in the literature across various sectors of physics \cite{colladay1998lorentz, kostelecky2002signals, kostelecky2004gravity, kostelecky2009electrodynamics, bailey2015short, kostelecky2016testing, kostelecky1989phenomenological, kostelecky1991photon, kostelecky1989spontaneous}, with the general aim of establishing stringent constraints that contribute to the search for a unified theory of quantum mechanics and gravity.

Investigations into LV have been conducted across diverse areas of physics, including scattering processes \cite{santos2017lorentz, cabral2024e, santos2018lorentz, cabral2023violation, cabral2024lorentz, casana2012effects, PhysRevD88025005, Charneski2012py, cabral2024exploring}, modified theories of gravity \cite{santos2015godel, jesus2019ricci, jesus2020godel, gama2017godel}, and vacuum fluctuation phenomena such as the Casimir effect \cite{ferreira2022tfd, ferreira2024thermo, correa2023aether}. In one of these works, Ref.~\cite{ferreira2022tfd}, the Casimir effect was investigated in the presence of a CPT-even aether-like vector term. In the present paper, we consider a symmetric traceless tensor motivated by investigations of LV effects in the kinetic sector of a massive scalar field \cite{altschul2006lorentz}, which lead to modifications in the propagator and induce anisotropic features in the corresponding Green function, thereby selecting preferred directions in the interaction \cite{cabral2024lorentz}.
 Moreover, the interplay between Lorentz violation and thermal effects has attracted increasing attention, especially within the TFD formalism. The combined treatment of LV and thermal corrections is crucial for understanding high-energy processes, particularly those believed to have occurred in the early universe.

This paper is organized as follows. In Section \ref{section2}, a brief overview of the fundamental aspects of TFD is provided. The definition of the energy-momentum tensor for the scalar field in the presence of a background Lorentz-violating tensor is presented in Section \ref{section3}. Section \ref{section4} is devoted to the Lorentz-violating scalar propagator, which plays a central role in the treatment of compactification effects. The main applications are examined in Section \ref{section5}, beginning with the calculation of the Stefan-Boltzmann-type law in Section \ref{section5a}, followed by the analysis of the Casimir effect at zero temperature in Section \ref{section5b}, and at finite temperature in Section \ref{section5c}. Finally, the concluding remarks are given in Section \ref{conclusion}.

\section{Thermo Field Dynamics}\label{section2}

In this section, the fundamental concepts of the Thermo Field Dynamics (TFD) formalism will be discussed, with an emphasis on how thermal and size effects can be incorporated through this approach. These elements are considered essential for the investigation of the Casimir effect at both zero and finite temperatures, as well as the Stefan-Boltzmann-type law.

TFD is a real-time formalism that incorporates temperature and the temporal evolution of a system using tools from quantum field theory \cite{Umezawa1}. In this approach, the statistical average is expressed as the expectation value of an arbitrary operator, which enables the definition of a thermal vacuum state. Two fundamental ingredients are required for this construction. The first ingredient is the duplication of the Hilbert space, denoted by $\mathcal{H}_T$, formed by the tensor product of the original space $\mathcal{H}$ and an auxiliary tilde space $\tilde{\mathcal{H}}$, such that $\mathcal{H}_T = \mathcal{H} \otimes \tilde{\mathcal{H}}$. An arbitrary operator, denoted by $Q$, satisfies the following conjugation rules within this formalism
\begin{eqnarray}
	&&\widetilde{(Q_iQ_j)}=\Tilde{Q}_i\Tilde{Q}_j;\\
	&&\widetilde{(cQ_i+Q_j)}=c^*\Tilde{Q}_i+\Tilde{Q}_j;\\
	&&\widetilde{(Q_i^\dagger)}=(\Tilde{Q}_i)^\dagger;\\
	&&\widetilde{(\Tilde{Q}_i)}=\xi Q_i,
\end{eqnarray}
where $\xi=-1$ for fermions and $\xi=+1$ for bosons. The second ingredient consists of the Bogoliubov transformations, which render the operator dependent on a new parameter known as the compactification parameter, defined as $\alpha = (\alpha_0, \alpha_1, \dots, \alpha_{D-1})$, where $D$ denotes the dimensionality of the system. The thermal effect is introduced by setting $\alpha_0 = \beta$ and $\alpha_1,\cdots\alpha_{D-1}=0$.  It is worth emphasizing that the TFD formalism can be interpreted from a topological perspective. In this context, the field theory is defined on the topology $\Gamma_D^\varrho = (\mathbb{S}^1)^\varrho \times \mathbb{R}^{D-\varrho}$, where $1 \leq \varrho \leq D$. Here, $D$ represents the total number of space-time dimensions, while $\varrho$ denotes the number of compactified ones. This framework allows for the compactification of any subset of the dimensions in the manifold $\mathbb{R}^D$, with each compactified dimension corresponding to a circle $\mathbb{S}^1$ of circumference $\alpha_n$, associated with the $n$-th direction.

Hence, as an application, the Bogoliubov transformations act on the operator $Q$ as follows
\begin{eqnarray}
	\begin{pmatrix}
		Q(\alpha)\\
		\tilde{Q}^\dagger(\alpha)
	\end{pmatrix}= \mathbb{M}(\alpha)\begin{pmatrix}
		Q\\
		\tilde{Q}^\dagger
	\end{pmatrix},\label{eq20}
\end{eqnarray}
where $\mathbb{M}(\alpha)$ is written, for fermions, as
\begin{eqnarray}
	\mathbb{M}_{F}(\alpha)=
	\begin{pmatrix}
		u(\alpha)& v(\alpha)\\
		-v(\alpha)& u(\alpha)
	\end{pmatrix},
\end{eqnarray}
with $u^2(\alpha)=1-f(\alpha)$ and $v^2(\alpha)=f(\alpha)$. Here, the function $f(\alpha)$ represents the Fermi-Dirac distribution, given by $f(\alpha)=\frac{1}{e^{\alpha \omega}+1}$.  For bosons, the Bogoliubov transformation becomes
\begin{eqnarray}
	\mathbb{M}_{B}(\alpha)=
	\begin{pmatrix}
		u^\prime(\alpha)& v^\prime(\alpha)\\
		v^\prime(\alpha)& u^\prime(\alpha)
	\end{pmatrix},
\end{eqnarray}
where $u^{\prime 2}(\alpha)=1+n(\alpha)$ and $v^{\prime 2}(\alpha)=n(\alpha)$, with $n(\alpha)=\frac{1}{e^{\alpha \omega}-1}$ being the Bose-Einstein distribution. Since scalar theory applications are being considered, the contribution $M_{B}(\alpha)$ will be taken into account.

In the TFD formalism, the scalar field propagator is defined as
\begin{align}
	G^{(ab)}(x-x^\prime; \alpha)=i\bra{0,\tilde{0}}\mathcal{T} [\phi^a(x;\alpha)\phi^{b}(x^\prime;\alpha)]\ket{0,\tilde{0}},
\end{align}
where $a,b=1,2$, denote the components of the thermal matrix in \eqref{eq20}, with $1(2)$ corresponding to the non-tilde (tilde) sector, ${\mathcal{T}}$ represents the time-ordering operator, and the field $\phi(x;\alpha)$ is defined by $\phi(x;\alpha)=M_{B}(\alpha)\phi(x)M_{B}(\alpha)^{-1}.$ In momentum space, the Green function is written as
\begin{align}
	{G}^{(ab)}(x-x^\prime;\alpha)=i\int\frac{d^4q}{(2\pi)^4}e^{-iq(x-x^\prime)}{G}^{(ab)}(q;\alpha).
\end{align}
It is important to note that, although the TFD formalism involves a doubling of the Hilbert space, the physical quantities depend only on the non-tilde sector, that is, $a=b=1$. With this in mind, the Green function is given by
\begin{align}
	{G}^{(11)}(q;\alpha)=G_0(q)+v^{\prime 2}(q;\alpha)\left[G^*_0(q)-G_0(q)\right],\label{propthermal}
\end{align}
where $G_0(q)$ is the standard scalar field propagator and $v^{\prime 2}$ is the generalized Bogoliubov transformation for compactified fields \cite{khanna2011quantum}, defined as
\begin{align}
	v^{\prime2}(q;\alpha)=\sum_{s=1}^\varrho\sum_{\{\sigma_s\}}2^{s-1}\sum_{l_{\sigma_1},\dots,l_{\sigma_s}=1}^\infty (-\xi)^{s+\sum_{r=1}^s l_{\sigma_r}}\exp\left[-\sum_{j=1}^s\alpha_{\sigma_j}l_{\sigma_j}q^{\sigma_j}\right],\label{BOGO}
\end{align}
with $\varrho$ representing the number of compactified dimensions, $\{\sigma_s\}$ being the set of all combinations with $s$ elements, and $q$ denoting the 4-momentum. This expression is consistent with the thermal formulation of compactified fields, and further details regarding its derivation can be found in Ref. \cite{Book}.

Once the fundamental concepts of TFD have been presented, the massive scalar field in the presence of a background field $k_{\mu\nu}$, which breaks Lorentz symmetry, will be considered. To investigate applications using TFD tools, the energy-momentum tensor associated with the theory will be calculated.

\section{The energy-momentum tensor}\label{section3}

In this section, the massive scalar field within a Lorentz-violating framework is considered. The Lagrangian describing this theory in the thermal representation of the TFD formalism is given by
\begin{eqnarray}
	\hat{{\cal L}}&=&{\cal L}-\tilde{\cal L}\nonumber\\
	&=&(g^{\mu\nu}+ k^{\mu\nu})\partial_\mu\phi\partial_\nu\phi-\frac{1}{2}m^2\phi^2-(g^{\mu\nu}+k^{\mu\nu})\partial_\mu\tilde{\phi}\partial_\nu\tilde{\phi}+\frac{1}{2}m^2\tilde{\phi^2},\label{FullL}
\end{eqnarray}
where $\mathcal{L}$ and $\tilde{\mathcal{L}}$ represent the usual and tilde spaces, respectively. Furthermore, the Lorentz-violating term $k^{\mu\nu}$ is a traceless symmetric tensor field belonging to the bosonic sector of the SME \cite{altschul2006lorentz}, which leads to the breaking of Lorentz symmetry. Since the measurable quantities are represented by the non-tilde variables of the Lagrangian in (\ref{FullL}), only elements from this space will be considered to obtain the energy-momentum tensor $T^{\mu\nu}$.

Considering the definition
\begin{eqnarray}
	{T^{\mu\nu}}=\frac{\partial \mathcal{L}}{\partial(\partial_\mu\phi)}\partial^\nu\phi-g^{\mu\nu}\mathcal{L}
\end{eqnarray}
and using only the non-tilde part of the Lagrangian, the energy-momentum is given by
\begin{eqnarray}
{T^{\mu\nu}}=\partial^\mu\phi\partial^\nu\phi+b^{\mu\rho}\partial_\rho\phi\partial^\nu\phi-\frac{1}{2}g^{\mu\nu}\partial_\alpha\phi\partial^\alpha\phi-\frac{1}{2}g^{\mu\nu}b^{\alpha\beta}\partial_\alpha\phi\partial_\beta\phi+\frac{1}{2}g^{\mu\nu}m^2\phi^2.
\end{eqnarray}

To avoid divergences arising from the product of two operators at the same space-time point, the energy-momentum tensor is expressed as
\begin{align}
	{T^{\mu\nu}}(x)&=\lim_{x'\to x}{\mathcal{T}}{\bigg\{}\partial^\mu\phi(x)\partial^\nu\phi(x')+{ k^{\mu\rho}}\partial_\rho\phi(x)\partial^\nu\phi(x')-\frac{1}{2}g^{\mu\nu}\partial_\alpha\phi(x)\partial^\alpha\phi(x')\nonumber\\&-\frac{1}{2}g^{\mu\nu}{k^{\alpha\beta}}\partial_\alpha\phi(x)\partial_\beta\phi(x')+\frac{1}{2}g^{\mu\nu}m^2\phi(x)\phi(x'){\bigg\}}.
\end{align}
In this way, the contact terms, which are the divergent contributions, are isolated from the finite and physically meaningful parts of the vacuum energy induced by the boundary conditions of the Casimir effect and by thermal effects. This procedure is analogous to the Casimir subtraction, thereby preserving the consistency of the formalism.

Using the canonical quantization of the scalar field
\begin{align}
	[\phi(x),\partial^\mu\phi(x')]=in^\mu_0\delta(\vec{x}-\vec{x}'),
\end{align}
with $n^\mu_0=(1,0,0,0)$ being a time-like vector and applying the time ordering to all terms, the energy-momentum tensor becomes
\begin{align}
	{T^{\mu\nu}}(x)&=\lim_{x'\to x}{\bigg\{}\Gamma^{\mu\nu}{ \mathcal{T}}[\phi(x)\phi(x')]+I^{\mu\nu}\delta(x-x'){\bigg\}}\label{6},
\end{align}
where 
\begin{align}
	\Gamma^{\mu\nu}=\partial^\mu\partial'^\nu+g_{\lambda\rho}{ k^{\mu\rho}}\partial^\lambda\partial'^\nu-\frac{1}{2}g_{\lambda\alpha}g^{\mu\nu}\partial^\gamma\partial'^\alpha-\frac{1}{2}g^{\mu\nu}g_{\lambda\alpha}g_{\rho\beta}{ k^{\alpha\beta}}\partial^\lambda\partial'^\rho+\frac{1}{2}g^{\mu\nu}m^2
\end{align}
and 
\begin{align}
	I^{\mu\nu}=in_0^\mu n^\nu_0+ig_{\lambda\rho}{ k^{\mu\rho}}-\frac{i}{2}g^{\mu\nu}g_{\gamma\alpha}n^\gamma_0n^\alpha_0-\frac{i}{2}g^{\mu\nu}g_{\lambda\alpha}g_{\rho\beta}{ k^{\alpha\beta}}n_0^\lambda n^\rho_0.
\end{align}

Taking the vacuum expectation value of Eq.~\eqref{6} leads to
\begin{align}
	\ev{{T^{\mu\nu}}(x)} &= \ev{{T^{\mu\nu}}(x)}{0} \nonumber\\
	&= \lim_{x' \to x} \left\{ \Gamma^{\mu\nu} \ev{{\mathcal{T}}[\phi(x)\phi(x')]}{0} + I^{\mu\nu} \delta(x - x') \right\}, \label{8}
\end{align}
where $\ev{{\mathcal{T}}[\phi(x)\phi(x')]}{0}$ represents the scalar field propagator, which can be defined as  
\begin{align}
	\ev{{\mathcal{T}}[\phi(x)\phi(x')]}{0} = iG_0(x - x').\label{PROPG}
\end{align}
It is important to note that, since the Lorentz-violating contribution is introduced in the kinematical part of the Lagrangian~\eqref{FullL}, the propagator of the massive scalar field is consequently modified. The Lorentz-violating propagator will be discussed in the next section.

The vacuum expectation value of the energy-momentum tensor in the TFD approach can now be written. Taking into account the duplication of the Hilbert space and the application of the Bogoliubov transformation, it is expressed as
\begin{align}
	\ev{{T^{\mu\nu}}(x;\alpha)} = \ev{{T^{\mu\nu}}(x)}{0(\alpha)} \label{10}.
\end{align}

To obtain a finite quantity, a renormalization procedure known as the Casimir prescription is introduced. As a consequence, the result is given by
\begin{align}
	{ \overline{T}^{\mu\nu(ab)}(x;\alpha) }&= \ev{{T^{\mu\nu(ab)}}(x;\alpha)} - \ev{{T^{\mu\nu(ab)}}(x)} \nonumber\\
	&= \lim_{x' \to x}\left\{\Gamma^{\mu\nu}(x,x') {\overline{G}^{(ab)}}(x - x';\alpha)\right\},\label{EMT}
\end{align}
where 
\begin{align}
	{\overline{G}^{(ab)}}(x - x';\alpha) = {G^{(ab)}}(x - x';\alpha) - G^{(ab)}_0(x - x').
\end{align}
{Here, the barred quantities denote renormalized quantities.}

Before proceeding with applications of the finite energy-momentum tensor, the propagator of the massive scalar field in the presence of the Lorentz-violating term will be calculated in the next section.

\section{The Lorentz-violating scalar propagator}\label{section4}

In this section, the general structure of the Lorentz-violating scalar propagator is addressed. Reference \cite{cabral2024lorentz} examined thermal effects on the Yukawa potential and scattering, describing the propagator in momentum space. { However, the approach followed here is based on the SME analyses of the Yukawa sector in Ref. \cite{altschul2006lorentz}. Although that work investigates the same Lorentz-violating operator considered here, it does not address thermal effects. In contrast, in the present study we incorporate temperature through the compactification parameter within the TFD formalism.} Moreover, to investigate phenomena such as the { Stefan-Boltzmann-type} law and the Casimir effect, it is necessary to express the propagator in coordinate space.

Let's start by writing the dispersion relation for the non-tilde scalar term of the Lagrangian in (\ref{FullL}), which is given by
\begin{eqnarray}
	\left(g_{\mu\nu}+{ k_{\mu\nu}}\right)p^\mu p^\nu=m_{\phi}^2.
\end{eqnarray}
Assuming $p^\mu=(E_p,\Vec{p})$, {the eigenenergies are obtained easily as
\begin{eqnarray}
	E_p=-\frac{k_{0j}p^j}{1+k_{00}}+\frac{1}{(1+k_{00})}\left[(k_{0j}p^j)^2+(\varepsilon_p^2-p^{i}k_{ij}p^{j})(1+k_{00})^2\right]^{\frac{1}{2}},\label{eq03}
\end{eqnarray}}
with $\varepsilon_p^2=(p^2+m_\phi^2)$.

Let's take a moment to discuss some considerations about the background field. When we analyze temporal components of Lorentz-violating terms, such as { $k_{\mu\nu}$}, we are considering non-standard time evolutions of the dynamical fields. However, such a consideration may be problematic when processes like scattering are analyzed. More precisely, the formulation of the asymptotic limit of the particle's free states must be carefully treated \cite{kosteleckycs}. Therefore, in order to avoid this issue, two possible treatments of the LV term can be adopted:
(i) the redefinition of the scalar field, or
(ii) setting { $k_{\mu 0} = k_{0 \mu} = 0$} \cite{cabral2024exploring}.
Here, we adopt the second approach. Hence, the dispersion relation becomes
\begin{eqnarray}
	E_p=\left(\varepsilon_p^2- p^{i}k_{ij}p^{j}\right)^2.\label{eq14}
\end{eqnarray}

Now, returning to the discussion of the propagator, in \cite{cabral2024lorentz}, the propagator in the presence of a background field { $k_{\mu\nu}$} was written as
\begin{eqnarray}
	{G(r)}=i\int\frac{d^4q}{(2\pi)^4}\frac{e^{-iqr}}{q\cdot \eta\cdot q-m_\phi^2},\label{eq01}
\end{eqnarray}
where  the dispersion relation is given by
\begin{eqnarray}
	q\cdot \eta\cdot q-m_\phi^2=E_q^2-q_0^2,\label{eq02}
\end{eqnarray}
with $q\cdot \eta\cdot q=q_\mu \eta^{\mu\nu}q_\nu$ where {$\eta^{\mu\nu}=g^{\mu\nu}+k_{\mu\nu}$.}

By substituting Eq.  (\ref{eq02}) into Eq. (\ref{eq01}), factorizing the denominator polynomials, and applying the residue theorem, we obtain
\begin{align}
	{G(r)}&=\int\frac{d^3q}{(2\pi)^3}\frac{e^{i\Vec{q}\cdot\Vec{r}}}{2E_q}\left[\Theta(t)e^{-iE_qt}+\Theta(-t)e^{iE_qt}\right]\equiv I_1(r,t)+I_2(r,t),\label{eq05}
\end{align}
where $\Theta(t)$ is the step function and $I_2(r,t)=I_1(r,-t)$. Since each term can be expressed in terms of the other, it suffices to analyze only $I_1(r,t)$. 

Using the relation in \eqref{eq14} and expanding it in a power series { to second order} in the Lorentz-violating contribution, we obtain
{\begin{eqnarray}
	I_1(r,t)=\frac{\Theta(t)}{2}\int\frac{d^3q}{(2\pi)^3}\left[\frac{1}{\varepsilon_q}+\frac{q^{j}k_{j\ell}q^\ell}{2\varepsilon_q^2}\left(it+\frac{1}{\varepsilon_q}\right)+\frac{(q^j k_{j\ell} q^\ell)^2}{8\varepsilon_q^3}\left(\frac{3}{\varepsilon^2}+\frac{3it}{\varepsilon}-t^2\right)\right]e^{i(\vec{q}\cdot\vec{r}-\varepsilon_qt)}.
\end{eqnarray}}

Now, using results from \cite{tabelaintegral}, along with properties of Bessel functions and some tensor analysis, the integral can be expressed {as
\begin{align}
	I_1(r,t)&=\frac{\Theta(t)}{4\pi^2}\Biggr\{\frac{m}{s}K_1(ms)-\frac{m^2r^2}{2s^2}K_2(ms)k^{ij}\hat{r}_i\hat{r}_j\nonumber\\&-\left[\frac{1}{4s^2}K_0(ms)-\frac{m^2s^2-2}{4ms^3}K_1(ms)-\frac{it}{32r^5}\mathbb{I}_1(r,t)\right]k^{j\ell}k_{j\ell}\nonumber\\&-\left[\frac{r^2(m^2s^2-4)}{2s^4}K_0(ms)+\frac{r^2(m^2s^2-8)}{2ms^5}K_1(ms)+\frac{it}{16r^5}\mathbb{I}_2(r,t)\right]k^{ij}k_{j\ell}\hat{r}_i\hat{r}^\ell\nonumber\\&+\left[\frac{3r^4(m^2s^2-8)}{8s^6}K_0(ms)+\frac{m^4s^4-48}{8ms^7}K_1(ms)+\frac{it}{64r^5}\mathbb{I}_3(r,t)\right]\left(k^{ij}\hat{r}_i\hat{r}_j\right)^2\Biggr\},
\end{align}}
with $K_1(x)$ denoting the modified Bessel function of the second kind, $s=\sqrt{r^2-t^2}$ representing the space-like interval, and $\hat{r}_i=x_i/r$ for $i=1,2,3$. { In addition, we have defined the following integrals, 
\begin{align}
\mathbb{I}_n(r,t)&=\left[1+(n-1)(n-2)\right]r^4 \int_0^\infty \frac{p^5 e^{-i t \epsilon }}{\epsilon ^3} \sin (p r)dp\nonumber\\&-\frac{(2n)!}{2^{n-1}n!} r^2 \int_0^\infty  \frac{p^3 e^{-i t
   \epsilon }}{\epsilon ^3}  \left[2 p r \cos (p r)+\left(p^2 r^2-2\right) \sin (p r)\right]dp\nonumber\\&+\frac{(2n+1)!}{3(2^n)n!}\int_0^\infty  \frac{p e^{-i t \epsilon }}{\epsilon ^3} \left[\left(p^4 r^4-12 p^2
   r^2+24\right) \sin (p r)-4 p r \left(p^2 r^2+6\right) \cos (p r)\right]dp
\end{align}
which, though seems divergent, can be treated in the distribution formalism.} Therefore, the complete expression for the Green function is given by {
{\small\begin{align}
	 G(r) &=\frac{1}{4\pi^2}\Biggr\{\frac{m}{s}K_1(ms)-\frac{m^2r^2}{2s^2}K_2(ms)k^{ij}\hat{r}_i\hat{r}_j-\left[\frac{1}{4s^4}K_0(ms)-\frac{m^2s^2-2}{4ms^3}K_1(ms)\right]k^{j\ell}k_{j\ell}\nonumber\\&-\left[\frac{r^2(m^2s^2-4)}{2s^4}K_0(ms)+\frac{r^2(m^2s^2-8)}{2ms^5}K_1(ms)\right]k^{ij}k_{j\ell}\hat{r}_i\hat{r}^\ell\nonumber\\&+\left[\frac{3r^4(m^2s^2-8)}{8s^6}K_0(ms)+\frac{m^4s^4-48}{8ms^7}K_1(ms)\right]\left(k^{ij}\hat{r}_i\hat{r}_j\right)^2\Biggr\}\nonumber\\
	&+\frac{i|t|}{256\pi^2r^5}\left[2\mathbb{I}_1(r,|t|)k^{j\ell}k_{j\ell}-4\mathbb{I}_2(r,|t|)k^{ij}k_{j\ell}\hat{r}_i\hat{r}^\ell+\mathbb{I}_3(r,|t|)\left(k^{ij}\hat{r}_i\hat{r}_j\right)^2\right]\label{PROP}.
\end{align}}}
Hence, Eq. \eqref{PROP} explicitly expresses the scalar field propagator in coordinate space, incorporating the effects of Lorentz violation through the background field {$k_{\mu\nu}$}. { Despite the presence of the $r^2$ or $s^2$ terms in the propagator, there is no problem in the large- $r$ regime within the $r\sim t$ limit, since for particles of this type we have light-like behavior ($m\sim0$), and the presence of Bessel functions ensures that the system remains well behaved.}

In the following sections, we explore the aforementioned processes at both zero and finite temperatures, highlighting the contributions of each component of the Lorentz-violating Green function to observable effects. These results will be contrasted with those found in the literature in the absence of Lorentz violation.

\section{Some applications: TFD and Lorentz violation}\label{section5}

In this section, the { Stefan-Boltzmann-type} law and the Casimir effect, at both zero and finite temperature, are investigated by considering different choices of the compactification parameter $\alpha$ within the framework of a Lorentz-violating massive scalar field theory.

\subsection{{Stefan-Boltzmann-type} law for a massive scalar field}\label{section5a}

To study {a Stefan-Boltzmann-type} law, the topology $\Gamma^1_4 = \mathbb{S}^1 \times \mathbb{R}^3$ is adopted. This choice implies that the compactification parameter is taken as $\alpha = (\beta, 0, 0, 0)$. In other words, within this topological structure, the time direction is compactified on a circle of length $\beta$. Consequently, the Bogoliubov transformation in Eq.~\eqref{BOGO} takes the form
\begin{align}
	v^{\prime2}(\beta)=\sum_{l_0=1}^{\infty}e^{-\beta q^0l_0}.
\end{align}
In this case, the Green function can be written as
\begin{align}
	{\overline{G}(x-x^\prime;\beta)}=2\sum_{l_0=1}^{\infty}{G(x-x^\prime-i\beta l_0 n_0)},
\end{align}
where $n_0^\mu=(1,0,0,0)$.

By incorporating these elements into the energy-momentum tensor given in Eq.~\eqref{EMT} and focusing on the component with $\mu = \nu = 0$, we obtain
{ {\small \begin{align}
	\overline{T}^{00(11)}(\beta)&=\lim_{x^\prime\to x}\sum_{l_0=1}^{\infty}\Big[m^2+\partial_0\partial^\prime_0+(1-k_{11})\partial_1\partial^\prime_1+(1-k_{22})\partial_2\partial^\prime_2+(1-k_{33})\partial_3\partial^\prime_3\nonumber\\&-k_{12}\left(\partial_1\partial_2^\prime+\partial_2\partial_1^\prime\right)-k_{13}\left(\partial_1\partial_3^\prime+\partial_3\partial_1^\prime\right)-k_{23}\left(\partial_2\partial_3^\prime+\partial_3\partial_2^\prime\right)\Big]\times G(x-x^\prime-i\beta l_0 n_0).\label{65}
\end{align}} }
Using the propagator given by Eq.~\eqref{PROP}, { we get an expression which represents the { Stefan-Boltzmann-type} law for a massive scalar field, incorporating corrections from the Lorentz-violating background tensor $k_{\mu\nu}$. By the nature and the quantity of the terms, the result is too large.}

To facilitate a clearer interpretation of the result and better understand the physical implications of the Lorentz-violating correction, we now consider the limit in which the scalar field is massless, allowing us to recover the standard { Stefan-Boltzmann-type} law and isolate the effects introduced solely by the Lorentz-violating term. Then Eq. (\ref{65}) becomes
{\begin{align}
	 E_A(\beta)\equiv \overline{T}^{00(11)}(\beta)=\frac{\pi^2}{30\beta^4}\left[1+\frac{1}{4} \left(k_{11}^2+k_{22}^2+k_{33}^2\right)+\frac{1}{2}\left(k_{12}^2+k_{13}^2+ k_{23}^2 \right)\right].\label{eq07}
\end{align}}
It is evident that the {Stefan-Boltzmann-type} law is directly affected by a positive contribution from the Lorentz-violating term, leading to an increase in the energy density. This behavior is particularly noteworthy for the specific background tensor considered here, as in other studies - such as Ref.~\cite{ferreira2022tfd} - {the Lorentz-violating effects arise from a different type of violation term, leading the resulting modifications to depend on the orientation or on specific components of the background field.}

For completeness, by examining the $\mu = \nu = 3$ component of the energy-momentum tensor, the pressure along the $z$-direction can be determined from {
{\small\begin{align}
\overline{T}^{33(11)}(\beta)&= \lim_
	{x^\prime\to x}\sum_{l_0=1}^{\infty}\biggl[\partial_0\partial^\prime_0+(k_{11}-1)\partial_1\partial^\prime_1+(k_{22}-1)\partial_2\partial^\prime_2-(k_{33}-1)\partial_3\partial^\prime_3\nonumber\\&+k_{12}(\partial_1\partial_2^\prime+\partial_2\partial_1^\prime)+k_{13}(\partial_3\partial_1^\prime-\partial_1\partial_3^\prime)+k_{23}(\partial_3\partial_2^\prime-\partial_2\partial_3^\prime)-m^2\biggr]G(x-x^\prime-i\beta l_0 n_0),
\end{align}}
leading to the expression of the pressure of the massive field depending on the temperature.

When the scalar field is massless, the result reduces to
\begin{eqnarray}
P_A(\beta)\equiv \overline{T}^{33(11)}(\beta)&=\frac{\pi^2}{90\beta^4}\left[1+\frac{1}{4} \left(k_{11}^2+k_{22}^2+k_{33}^2\right)+\frac{1}{2}\left(k_{12}^2+k_{13}^2+ k_{23}^2 \right)\right]\label{eq11}
\end{eqnarray}}
This reveals that the { both the energy and pressure are} directly affected not only by diagonal but also by non-diagonal components of the Lorentz-violating background. Depending on the values of the parameters that break the symmetry, these contributions can either enhance or suppress the pressure.

{ It should be emphasized that this result exhibits a Stefan-Boltzmann-type behavior for the thermal energy density inside the cavity. It does not correspond to a full Stefan-Boltzmann radiation law. In the SME framework, establishing such a radiation law would require an independent analysis of the energy flux, derived from the appropriate components of the energy-momentum tensor, which lies beyond the scope of the present work.}

\subsection{Casimir effect at zero temperature}\label{section5b}

In what follows, the Casimir effect at zero temperature will be examined in the presence of the background field { $k_{\mu\nu}$}. For this analysis, the compactification parameter is taken as $\alpha = (0, 0, 0, 2i d)$, as this configuration allows the topology to be compactified along the $z$-axis. This choice is considered appropriate, given that the theory is defined on the topology $\Gamma^1_4 = \mathbb{S}^1 \times \mathbb{R}^3$, where $\mathbb{S}^1$ represents a circle of length $L = 2d$. This representation allows the Bogoliubov transformation to be written as 
\begin{align}
	v^{\prime2}(d)=\sum_{l_3=1}^\infty e^{-i2dq^3l_3}.
\end{align}
Then, the Green function is obtained in the form
\begin{align}
	{\overline{G}(x-x^\prime)}=2\sum_{l_3=1}^\infty {G(x-x^\prime-2dl_3 n_3)}.
\end{align}

In order to calculate the Casimir energy, the component $\mu = \nu = 0$ of the energy-momentum tensor is considered, which results in
{ {\small \begin{align}
	\overline{T}^{00(11)}(d)&=\lim_{x^\prime\to x}\sum_{l_0=1}^{\infty}\Big[m^2+\partial_0\partial^\prime_0+(1-k_{11})\partial_1\partial^\prime_1+(1-k_{22})\partial_2\partial^\prime_2+(1-k_{33})\partial_3\partial^\prime_3\nonumber\\&-k_{12}\left(\partial_1\partial_2^\prime+\partial_2\partial_1^\prime\right)-k_{13}\left(\partial_1\partial_3^\prime+\partial_3\partial_1^\prime\right)-k_{23}\left(\partial_2\partial_3^\prime+\partial_3\partial_2^\prime\right)\Big]\times G(x-x^\prime-2dl_3 n_3).
	\end{align}}}   It is worth noting that this expression shares the same structure as the {Stefan-Boltzmann-type} law, except that the compactification parameter now acts on a different component of the propagator. By substituting the modified propagator into the previously obtained expression for the energy-momentum tensor, {the Casimir energy is obtained. }

For simplicity, the limit $m \to 0$ is considered. In this case, the Casimir energy for a massless scalar field in the presence of the Lorentz-violating background field { $k_{\mu\nu}$ is given by
\begin{align}
	E_B(d)\equiv\overline{T}^{00(11)}(d)=-\frac{\pi^2}{1440d^4}\left[1+\frac{1}{4}\left(k_{11}^2+k_{22}^2+5k_{33}^2\right)+\frac{1}{2}\left(k_{12}^2-3k_{13}^2-3k_{23}^2\right)-\frac{5}{4}k_{33}\right].\label{eq08}
\end{align} }
The Casimir energy exhibits explicit corrections arising from the Lorentz-violating background coefficients {$k_{\mu\nu}$}. In particular, the terms { of the LV contribution including the $3-$ direction} dominate the modifications along the direction perpendicular to the plates, contributing additively and thus enhancing the attractive force between them. These alterations reveal anisotropic vacuum fluctuations induced by the Lorentz violation, which could potentially be probed in future experimental investigations.

By analogy, the Casimir pressure can be obtained by setting $\mu = \nu = 3$. Consequently, the corresponding component of the energy-momentum tensor is given by
{ {\small\begin{align}
	\overline{T}^{33(11)}(\beta)&= \lim_
	{x^\prime\to x}\sum_{l_0=1}^{\infty}\biggl[\partial_0\partial^\prime_0+(k_{11}-1)\partial_1\partial^\prime_1+(k_{22}-1)\partial_2\partial^\prime_2-(k_{33}-1)\partial_3\partial^\prime_3\nonumber\\&+k_{12}(\partial_1\partial_2^\prime+\partial_2\partial_1^\prime)+k_{13}(\partial_3\partial_1^\prime-\partial_1\partial_3^\prime)+k_{23}(\partial_3\partial_2^\prime-\partial_2\partial_3^\prime)-m^2\biggr]G(x-x^\prime-2dl_3 n_3),
\end{align}}}
which leads to the Casimir pressure, { in the massless limit $m\to0$, given by  
\begin{align}
	P_B(d)\equiv\overline{T}^{33(11)}(d)=-\frac{\pi^2}{480d^4}\left[1+\frac{1}{4}\left(k_{11}^2+k_{22}^2+5k_{33}^2\right)+\frac{1}{2}\left(k_{12}^2-3k_{13}^2-3k_{23}^2\right)-\frac{5}{4}k_{33}\right].\label{eq12}
\end{align}}
The results above describes the Casimir { energy and pressure} at zero temperature in the presence of Lorentz-violating contributions. In this case, we notice that there is a linear and quadratic dependence on the background field. The linear contribution arises from the { isolated component $k_{33}$, that is, the transverse direction. In addition, the quadratic terms are the dominant ones since that $k_{\mu\nu}$ lies at a perturbative level}. Therefore, depending on the specific values of $k_{\mu\nu}$, the Casimir pressure may either increase or decrease, demonstrating that Lorentz violation alters the vacuum quantum corrections associated with the scalar field.

\subsection{Casimir effect at finite temperature}\label{section5c}

The main focus of this section is to analyze the Casimir effect at finite temperature. In this framework, the compactification parameter will account not only for the spatial component along the $z$-axis, which encodes the system's finite size, but also for the temporal component, representing thermal effects. Specifically, we consider $\alpha = (\beta, 0, 0, i2d)$, corresponding to the topology $\Gamma^2_4=\mathbb{S}^1\times\mathbb{S}^1\times\mathbb{R}^2$. Consequently, the Bogoliubov transformation takes the form:
\begin{align}
	v^{\prime2}(\beta,d)=\sum_{l_0=1}^\infty e^{-\beta q^0l_0}+\sum_{l_3=1}^\infty e^{-2idq^3l_3}+\sum_{l_0,l_3=1}^\infty e^{-\beta q^0 l_0-2idq^3l_3}.\label{eq06}
\end{align}
It can be observed that the first term in Eq.\eqref{eq06} corresponds to the calculation presented in Section \ref{section5a}, while the second term represents the Casimir effect discussed in Section~\ref{section5b}. Both of these contributions have already been addressed. Therefore, the latter term in Eq.~\eqref{eq06} will be examined in greater detail here. Accordingly, the Green function associated with this contribution takes the form
\begin{align}
	{\overline{G}(x-x^\prime;\beta,d)}=4\sum_{l_0,l_3=1}^\infty {G(x-x^\prime-i\beta l_0 n_0-2dl_3n_3)}.
\end{align}
In this topological structure, the Casimir energy { at finite temperature, when the scalar field is considered massless, i.e., in the limit $m\to0$, is given by
\begin{align}
\overline{T}^{00(11)}(\beta,d)&=\frac{1}{\pi^2}\sum_{l_0,l_3=1}^\infty\Bigg\{\frac{9 \pi  (\beta l_0)  \left[2 \left(k_{11}^2+2 k_{12}^2-8 k_{13}^2+k_{22}^2-8 k_{23}^2\right)-3
   k_{33}^2\right] }{1024 d^5 l_3^5}\nonumber\\&-\frac{32 (dl_3)^6 \left[k_{11}^2+2 k_{12}^2-6
   k_{13}^2+k_{22}^2-6 k_{23}^2+k_{33} \left(5 k_{33}-8\right)+4\right]}{\left(4 d^2 l_3^2+\beta ^2 l_0^2\right){}^5}\nonumber\\&+\frac{7 (\beta l_0)^2 (dl_3)^4
   \left[k_{11}^2+2 k_{12}^2-30 k_{13}^2+k_{22}^2-30 k_{23}^2+k_{33} \left(53 k_{33}-32\right)+4\right]}{\left(4 d^2 l_3^2+\beta ^2 l_0^2\right){}^5}\nonumber\\&+\frac{10 (\beta l_0)^4 (dl_3)^2 \left[k_{11}^2+2 k_{12}^2-6 k_{13}^2+k_{22}^2-6 k_{23}^2-k_{33} \left(7
   k_{33}+8\right)+4\right]}{\left(4 d^2 l_3^2+\beta ^2 l_0^2\right){}^5}\nonumber\\&+\frac{3 (\beta l_0)^6 \left(k_{11}^2+2 k_{12}^2+2 k_{13}^2+k_{22}^2+2
   k_{23}^2+k_{33}^2+4\right) }{2\left(4 d^2 l_3^2+\beta ^2 l_0^2\right){}^5}\Bigg\}.\label{eq09}
\end{align}}

To properly determine the total energy ${E_\text{tot}}(\beta,d)$ of the system, all contributions from the Bogoliubov transformation in Eq.~\eqref{eq06} must be taken into account. This means that, in addition to Eq.\eqref{eq09}, we must also revisit Eqs.~\eqref{eq07} and \eqref{eq08}, leading to
{\begin{align}
E_\text{tot}(\beta,d)&=\frac{1}{\pi^2}\sum_{l_0,l_3=1}^\infty\Bigg\{\frac{9 \pi  (\beta l_0)  \left[2 \left(k_{11}^2+2 k_{12}^2-8 k_{13}^2+k_{22}^2-8 k_{23}^2\right)-3
   k_{33}^2\right] }{1024 d^5 l_3^5}\nonumber\\&-\frac{32 (dl_3)^6 \left[k_{11}^2+2 k_{12}^2-6
   k_{13}^2+k_{22}^2-6 k_{23}^2+k_{33} \left(5 k_{33}-8\right)+4\right]}{\left(4 d^2 l_3^2+\beta ^2 l_0^2\right){}^5}\nonumber\\&+\frac{7 (\beta l_0)^2 (dl_3)^4
   \left[k_{11}^2+2 k_{12}^2-30 k_{13}^2+k_{22}^2-30 k_{23}^2+k_{33} \left(53 k_{33}-32\right)+4\right]}{\left(4 d^2 l_3^2+\beta ^2 l_0^2\right){}^5}\nonumber\\&+\frac{10 (\beta l_0)^4 (dl_3)^2 \left[k_{11}^2+2 k_{12}^2-6 k_{13}^2+k_{22}^2-6 k_{23}^2-k_{33} \left(7
   k_{33}+8\right)+4\right]}{\left(4 d^2 l_3^2+\beta ^2 l_0^2\right){}^5}\nonumber\\&+\frac{3 (\beta l_0)^6 \left(k_{11}^2+2 k_{12}^2+2 k_{13}^2+k_{22}^2+2
   k_{23}^2+k_{33}^2+4\right) }{2\left(4 d^2 l_3^2+\beta ^2 l_0^2\right){}^5}\Bigg\}+E_A(\beta)+E_B(d).\label{eq15}
\end{align}}

{The total energy presents a very complex dependence on the compactification parameters, in such a way that the features can be viewed separately. That is, first considering the thermal Casimir energy at fixed plate separations.} At short distances, the interaction between the plates can be attractive, repulsive, or even null, depending on the Lorentz-violating parameter and the temperature. As the temperature decreases (i.e., $\beta \to \infty$), the energy approaches the asymptotic behavior given by Eq.~\eqref{eq08}, as expected. However, when the temperature increases ($\beta \to 0$), the energy shows a pronounced {$\beta^{-4}$} dependence, characteristic of the { Stefan-Boltzmann-type} law, and the repulsive interaction becomes increasingly dominant. { On the other hand one can consider the behavior of the Casimir energy as a function of the distance between the plates for fixed temperatures.} At large separations, the energy is predominantly governed by thermal effects. In contrast, as the plates approach each other, thermal contributions become negligible, and the energy behavior transitions to that described by Eq.~\eqref{eq09}. Furthermore, it is important to note that all the analyses carried out so far are strongly dependent on the configuration of the Lorentz-violating parameters $k_{ij}$, which play a significant role in determining whether the scalar field energy is increased, decreased, or exactly zero.

For the Casimir pressure at finite temperature, we set $\mu = \nu = 3$, and in the massless limit, we obtain
{ \begin{align}
	\overline{T}^{33(11)}(\beta,d)&=\frac{1}{\pi^2}\sum_{l_0,l_3=1}^\infty\Bigg\{-\frac{9 \pi  (\beta l_0)  \left[10k_{11}^2+20 k_{12}^2-48 k_{13}^2+10k_{22}^2-48 k_{23}^2+17
   k_{33}^2\right] }{1024 d^5 l_3^5}\nonumber\\&-\frac{96 (dl_3)^6 \left[k_{11}^2+2 k_{12}^2-6
   k_{13}^2+k_{22}^2-6 k_{23}^2+k_{33} \left(5 k_{33}-8\right)+4\right]}{\left(4 d^2 l_3^2+\beta ^2 l_0^2\right){}^5}\nonumber\\&-\frac{40 (\beta l_0)^2 (dl_3)^4
   \left[k_{11}^2+2 k_{12}^2+2 k_{13}^2+k_{22}^2 +2 k_{23}^2-11k_{33}+4\right]}{\left(4 d^2 l_3^2+\beta ^2 l_0^2\right){}^5}\nonumber\\&-\frac{2 (\beta l_0)^4 (dl_3)^2 \left[k_{11}^2+2 k_{12}^2+26 k_{13}^2+k_{22}^2+26 k_{23}^2+k_{33} \left(25
   k_{33}+24\right)+4\right]}{\left(4 d^2 l_3^2+\beta ^2 l_0^2\right){}^5}\nonumber\\&+\frac{(\beta l_0)^6 \left(k_{11}^2+2 k_{12}^2+2 k_{13}^2+k_{22}^2+2
   k_{23}^2+k_{33}^2+4\right) }{2\left(4 d^2 l_3^2+\beta ^2 l_0^2\right){}^5}\Bigg\}.\label{eq13}
\end{align}}

Taking into account Eqs.~\eqref{eq11}, \eqref{eq12}, and \eqref{eq13}, the total pressure is given by
{ \begin{align}
 P_\text{tot}(\beta,d)&=\frac{1}{\pi^2}\sum_{l_0,l_3=1}^\infty\Bigg\{-\frac{9 \pi  (\beta l_0)  \left[10k_{11}^2+20 k_{12}^2-48 k_{13}^2+10k_{22}^2-48 k_{23}^2+17
   k_{33}^2\right] }{1024 d^5 l_3^5}\nonumber\\&-\frac{96 (dl_3)^6 \left[k_{11}^2+2 k_{12}^2-6
   k_{13}^2+k_{22}^2-6 k_{23}^2+k_{33} \left(5 k_{33}-8\right)+4fan\right]}{\left(4 d^2 l_3^2+\beta ^2 l_0^2\right){}^5}\nonumber\\&-\frac{40 (\beta l_0)^2 (dl_3)^4
   \left[k_{11}^2+2 k_{12}^2+2 k_{13}^2+k_{22}^2 +2 k_{23}^2-11k_{33}+4\right]}{\left(4 d^2 l_3^2+\beta ^2 l_0^2\right){}^5}\nonumber\\&-\frac{2 (\beta l_0)^4 (dl_3)^2 \left[k_{11}^2+2 k_{12}^2+26 k_{13}^2+k_{22}^2+26 k_{23}^2+k_{33} \left(25
   k_{33}+24\right)+4\right]}{\left(4 d^2 l_3^2+\beta ^2 l_0^2\right){}^5}\nonumber\\&+\frac{(\beta l_0)^6 \left(k_{11}^2+2 k_{12}^2+2 k_{13}^2+k_{22}^2+2
   k_{23}^2+k_{33}^2+4\right) }{2\left(4 d^2 l_3^2+\beta ^2 l_0^2\right){}^5}\Bigg\}+P_A(\beta)+P_B(d).
\end{align}}

{The total Casimir pressure at finite temperature can be discussed in an analogous way as those played by the energy. First, one can analyse the influence of thermal effects at fixed values of $d$.} For plates that are close together, when the temperature is negligible, the total pressure approaches the asymptotic values given by Eq. \eqref{eq12}, as expected. As the temperature increases, the magnitude of the interaction can become strongly attractive or repulsive, a behavior determined exclusively by the Lorentz-violating parameters. { In a diametrically opposite way, when examining the dependence on distance for fixed temperatures, one can confirm the large-distance asymptotic behavior described by Eq. \eqref{eq11}.} Once again, the parameters $k_{ij}$ play a crucial role in determining both the nature (attractive or repulsive) and the strength of the Casimir pressure.

\section{Conclusion}\label{conclusion}

The fluctuations of a Lorentz-violating scalar field under the effects of space-time compactifications were investigated. Lorentz symmetry breaking was introduced through a background traceless antisymmetric tensor, while the topological modifications arose from the TFD formalism. In this context, new quantities emerged, including a modified stress-energy tensor and an altered Feynman propagator. The compactification parameters introduced temperature via the compactification of the temporal dimension, whereas the spatial compactification modeled the presence of two parallel plates. All results {can be} expressed in terms of the field mass $m$, although, for simplicity and clarity, the analysis was primarily conducted in the massless limit.

The energy-momentum tensor characterizes the energy density and pressure associated with the field fluctuations, leading to Lorentz-violating modifications of the {Stefan-Boltzmann-type} law and the Casimir effect. These modifications are significant and reveal novel behaviors. When both temporal and spatial compactifications are combined, the characteristic features of the thermal Casimir effect emerge. The results indicate that the compactification parameters $d$ and $T$ act as ``magnitude regulators'', controlling the prominence of these effects. Meanwhile, Lorentz violation governs the qualitative nature of these phenomena, notably influencing whether quantities such as energy and pressure become positive, negative, or vanish in regions where they would otherwise exhibit different behavior.

\section*{Acknowledgments}

This work by A. F. S. is partially supported by National Council for Scientific and Technological
Development - CNPq project No. 312406/2023-1. D. S. C., L. A. S. E. and L. H. A. R. thanks CAPES for financial support.

\section*{Data Availability Statement}

%No data are available because of the nature of the research. This publication is theoretical work that does not require supporting research data.
No Data associated in the manuscript.

\section*{Conflicts of Interest}

No conflict of interests in this paper.

%%%%%%%%%%%%%%%%%%%%%%%%%%%%%%%%%%%%%%%%%%%%%%%%%%%%%%%%%%%%%%%%%%%%%%%%%%%%%%%%%%%%%%%%%%%%%%%%%%%%%%%%%%%%%%%%%

\global\long\def\link#1#2{\href{http://eudml.org/#1}{#2}}
 \global\long\def\doi#1#2{\href{http://dx.doi.org/#1}{#2}}
 \global\long\def\arXiv#1#2{\href{http://arxiv.org/abs/#1}{arXiv:#1 [#2]}}
 \global\long\def\arXivOld#1{\href{http://arxiv.org/abs/#1}{arXiv:#1}}

%%%%%%%%%%%%%%%%%%%%%%%%%%%%%%%%%%%%%%%%%%%%%%%%%%%%%%%%%%%%%%%%%%%%%%%%%%%%%%%%%%%%%%%%%%%%%%%%%%%%%%%%%%%

\end{document}